\documentclass[aps,prl,twocolumn,showpacs,apsrev,superscriptaddress]{revtex4-1}
\usepackage[T1]{fontenc}
\usepackage{lmodern}
\usepackage[english]{babel}
\usepackage{amsmath}
\usepackage{amsfonts}
\usepackage{graphicx} 
\usepackage{comment}
\usepackage{xfrac}
\usepackage{color,soul}
\usepackage{multirow}
\usepackage{hyperref}
\usepackage{braket}
\usepackage{pstricks}
 \usepackage[normalem]{ulem}
 
\hypersetup{
	colorlinks = true,
	pdftitle = {},
	pdfauthor = {},
	pdfkeywords = {},
	linkcolor = blue,
	citecolor = blue,
	filecolor = black,
	urlcolor = magenta
}

\graphicspath{{./graphics/}}

\begin{document}

\title{Curved Magnetism in CrI$_3$ }
 
\author{Alexander Edstr\"om}
\affiliation{Institut  de  Ci\`encia  de  Materials  de  Barcelona  (ICMAB-CSIC),  Campus  UAB,  08193  Bellaterra, Spain}
\affiliation{Department of Applied Physics, School of Engineering Sciences,KTH Royal Institute of Technology, AlbaNova University Center, 10691 Stockholm, Sweden}
\author{Danila Amoroso}
\affiliation{Consiglio Nazionale delle Ricerche CNR-SPIN, c/o Universit\`{a} degli Studi 'G. D'Annunzio', 66100, Chieti, Italy}
\affiliation{NanoMat/Q-mat/CESAM, Universit\'e de Li\`ege, B-4000 Liege, Belgium}
\author{Silvia Picozzi}
\affiliation{Consiglio Nazionale delle Ricerche CNR-SPIN, c/o Universit\`{a} degli Studi 'G. D'Annunzio', 66100, Chieti, Italy}
\author{Paolo Barone}
\affiliation{Consiglio Nazionale delle Ricerche CNR-SPIN,  Area della Ricerca di Tor Vergata,Via del Fosso del Cavaliere 100, I-00133 Rome, Italy}
\author{Massimiliano Stengel}
\affiliation{Institut  de  Ci\`encia  de  Materials  de  Barcelona  (ICMAB-CSIC),  Campus  UAB,  08193  Bellaterra, Spain}
\affiliation{ICREA  -  Instituci\`{o}  Catalana  de  Recerca  i  Estudis  Avan\c{c}ats,  08010  Barcelona,  Spain}


\begin{abstract}
Curved magnets attract considerable interest for their unusually
rich phase diagram, often encompassing exotic (e.g., topological or chiral) spin states.
Micromagnetic simulations are playing a central role in the theoretical 
understanding of such phenomena; their predictive power, however, rests on the 
availability of reliable model parameters to describe a given material or nanostructure.
Here we demonstrate %
how non-collinear-spin polarized density-functional theory can be 
used to determine the flexomagnetic coupling coefficients in real systems.
By focusing on monolayer CrI$_3$, we find a crossover as a function of curvature between a 
magnetization normal to the surface to a cycloidal state, 
which we rationalize in terms of effective anisotropy and 
Dzyaloshinskii-Moriya contributions to the magnetic energy.
Our results reveal an unexpectedly large impact of spin-orbit interactions on the curvature-induced 
anisotropy, which we discuss in the context of existing phenomenological models.
\end{abstract}

\maketitle

\emph{Introduction - }  %
Inhomogeneous deformations in the form of local curvature are ubiquitous at the nanoscale, and currently regarded as a rich playground for new phenomena.~\cite{Flexoel_Zubkoetal,Streubel_2016,Sheka2021}  
Understanding their effects is crucial for the materials design of 
tailored functionalities, e.g., in flexible electronics, as well as for the 
tunability of existing ones via external mechanical stimuli~\cite{Lu59,Bhaskar2016}. 
A notable example is flexoelectricity~\cite{Flexoel_Zubkoetal}, i.e. the emergence of a macroscopic electric polarization in presence of a non-uniform strain, highly attractive for piezoelectric replacements~\cite{Bhaskar2016}  or  strain-enabled photovoltaics~\cite{Shu2020}.
Atomically thin two-dimensional (2D) crystals and membranes, due to their extreme flexibility and natural tendency towards rippling~\cite{Fasolino2007}, appear as the ideal class of materials to explore these effects.

Strain gradients can have a comparably strong impact on magnetism~\cite{Streubel_2016,PhysRevLett.123.077201,Sheka2021},
via a curvature-induced modification of the spin coupling parameters,  commonly referred to as \emph{flexomagnetism}~\cite{hertel}.
Representative examples include the emergence of curvature-induced Rashba spin-orbit coupling (SOC)~\cite{gentile1,gentile2,PYATAKOV2015255}, asymmetric magnon dispersions~\cite{PhysRevLett.117.227203}, topological magnetism~\cite{PhysRevLett.114.197204}, magnetic anisotropies and effective Dzyaloshinskii-Moriya interactions (DMI)~\cite{PhysRevLett.114.197204,PhysRevLett.112.257203,Sheka_2015,Streubel_2016}.
Remarkably, the geometric DMI can drive the formation of chiral and topological spin configurations even in absence of SOC~\cite{Streubel_2016}, thus lifting the traditional requirement of heavy elements in the crystal 
structure for exotic magnetic orders to occur.
An exciting development in this context is the recent experimental report of strain-gradient-induced DMI 
resulting in a room-temperature Skyrmion lattice~\cite{PhysRevLett.127.117204}.
Thanks to impressive advances in experimental fabrication and characterization techniques,~\cite{Pacheco2017,Donnelly2017} 
additional observations of these effects in the lab are anticipated in the near future.
From theory, it would be desirable to support the experimental efforts by developing a quantitatively accurate understanding of flexomagnetism in real materials. 
Several simplifying assumptions are currently adopted in micromagnetic simulations of curved nanostructures 
(see, e.g., Ref.~\cite{Streubel_2016} and references therein), potentially limiting their predictive power. 
For example, the effective DMI and anisotropy in the curved structure are modeled as non-relativistic effects 
via a coordinate transformation operated on the isotropic exchange interaction.
While appearing reasonable, such an approximation has not been tested in a realistic context, and its 
validity is still an open question.

\emph{Ab initio} electronic-structure methods
have played a leading role in understanding low-dimensional magnets
and formulating new predictions~\cite{HEIDE20092678,Xu2018,NiI2_us,Tiwari2021}. Their application to flexomagnetism is, however, still at an infancy stage and 
  technically challenging in the framework of density-functional theory (DFT): 
  Curvature breaks translational symmetry and its study requires, in principle,
  large periodic supercells that may contain several hundreds of atoms.
The recent discovery of long-range magnetic order in monolayers of CrI$_3$ and other Van der Waals compounds \cite{Huang2017,CGT_gong,mcguire,Burch2018} provides a natural playground to study curvature-induced effects on magnetic properties from first principles.
In addition to their practical~\cite{Song1214} and fundamental~\cite{Han2019,dong,NiI2_us} interest,
two-dimensional magnets with few atoms per surface unit allow for the simulation of bent geometries to be tractable (even if expensive) within DFT~\cite{KUKLIN2020114205}.
Nonetheless, the few existing studies have targeted collinear spin structures~\cite{Shen2018,Shi2019,KUKLIN2020114205}.  This is clearly insufficient for understanding the emergence of nontrivial magnetic states in a bent layer, which requires a fully non-collinear treatment of the spins in presence of curvature.

Here, we use non-collinear-spin DFT to investigate the magnetic properties 
of CrI$_3$ as a function of curvature.
The latter is incorporated by focusing on nanotube (NT) geometries 
with radii ($R$)  between 7.5~{\AA} and 30~\AA.
By comparing the energies of different magnetic states as functions of $R$, 
we show that curvature leads to a crossover 
between an out-of-plane 
magnetization for the flat monolayer (corresponding to a radial magnetization for large-$R$ NTs), and a cycloidal state   
at larger curvatures (smaller $R$), 
which is stabilized by an effective curvature-induced DMI~\cite{Sheka_2015}.
To rationalize this finding, we construct a continuum model, whose 
parameters are fully determined from first principles, in terms
of the spin stiffness, anisotropy and DMI strength and their dependence
on curvature.
We find that SOC,  which largely originates from the I atoms in CrI$_3$,~\cite{Lado_2017} has a surprisingly strong impact on  
the curvature-induced effective anisotropy, which qualitatively departs from the assumptions of earlier models.

\begin{figure}
	\centering
	\includegraphics[width=0.5\textwidth]{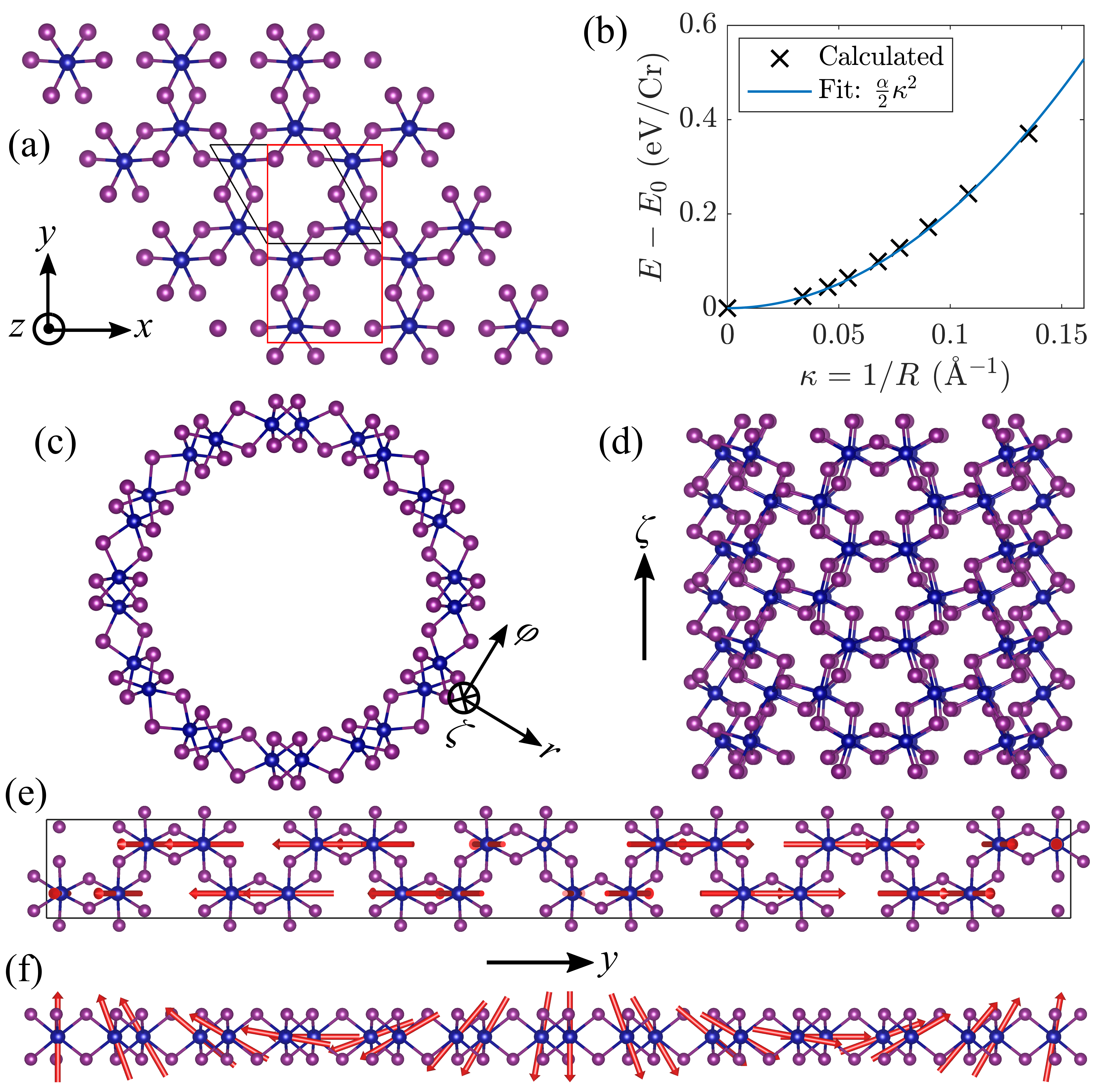}
	\caption{(a) Monolayer CrI$_3$ (Cr - blue, I -purple) with black lines showing the primitive 8 atom unit cell (2 Cr and 6 I) while the red lines show the doubled unit cell, $N$ of which are used to construct $(N,N)$ NTs. (b) Calculated energy (per Cr) as a function of curvature. (c) $N=6$ NT from above and (d) side. (e) Supercell strip for a spin spiral with wavelength equal to the circumference of the $N=6$ NT, seen from above and (f) side, with arrows showing the magnetic moments $\mathbf{m}_i = \left[ 0, -\sin(q y_i) , \cos(q y_i) \right]$. %
	}
	\label{fig.CrI3}
\end{figure}

\emph{Methods - }
Calculations are performed using the projector augmented wave (PAW)~\cite{PhysRevB.50.17953,PhysRevB.59.1758} method, as implemented in VASP~\cite{KRESSE199615,PhysRevB.49.14251,PhysRevB.47.558}. The local density approximation (LDA) is used for the exchange-correlation together with an additional Coulomb repulsion~\cite{PhysRevB.57.1505} of $U=0.5~\mathrm{eV}$ on  Cr d-states. The cutoff energy for the plane-waves basis set is 350~eV.~\cite{CompSettings} 
Calculations are performed within periodic boundary conditions, with (at least) {15~\AA} of vacuum separating the repeated images of the monolayers or NTs. 
A $4 \times 4 \times 1$ $k$-point mesh is used for calculations on the freestanding monolayer~\cite{kpts}, while  $4 \times 1 \times 1$ $k$-points are used for the NTs. The NT structures are optimized 
considering collinear ferromagnetism until forces are smaller than 3~meV/\AA. 
We consider $(N,N)$ armchair NTs~\cite{DRESSELHAUS1995883} %
for $N=4, 5, 6, 7, 8, 10, 12, 16$, as illustrated in Fig.~\ref{fig.CrI3}. This means that $N$ units of the cell marked with a red rectangle in Fig.~\ref{fig.CrI3}(a) are wrapped around the circumference.  The Cr-Cr distance is $d=3.87~\mathrm{\AA}$, whereby the NT radii are approximately $NR_0$ with $R_0=\frac{3d}{2\pi}=1.85~\AA$.
Energy calculations on different magnetic states were performed via constrained non-collinear magnetic calculations, with and without SOC, using a penalty energy for spins deviating from the desired configuration~\cite{PhysRevB.62.11556,PhysRevB.91.054420,suppl}. 

\emph{Results - }
 Energy differences of the relaxed NTs with collinear ferromagnetism, relative to the flat monolayer, are reported as a function of curvature $\kappa=1/R$ in Fig.~\ref{fig.CrI3}(b). 
 The calculated data show a smooth monotonic behavior with $\kappa$, confirming the accuracy of the structural relaxations.
 The fitted bending modulus, $\alpha = 2.2~\mathrm{eV}$, is similar as other monolayers of transition metal dichalcogenides~\cite{Kumar_2020}. 
This elastic contribution, of order 1 eV per Cr atom, dominates by far the energetics of bending. The dependence on the magnetic ordering is typically three orders of magnitude smaller ($\sim$1 meV/Cr), as we shall see in the following while
discussing our results with non-collinear spins and SOC.

\begin{figure}
	\centering
	\includegraphics[width=0.5\textwidth]{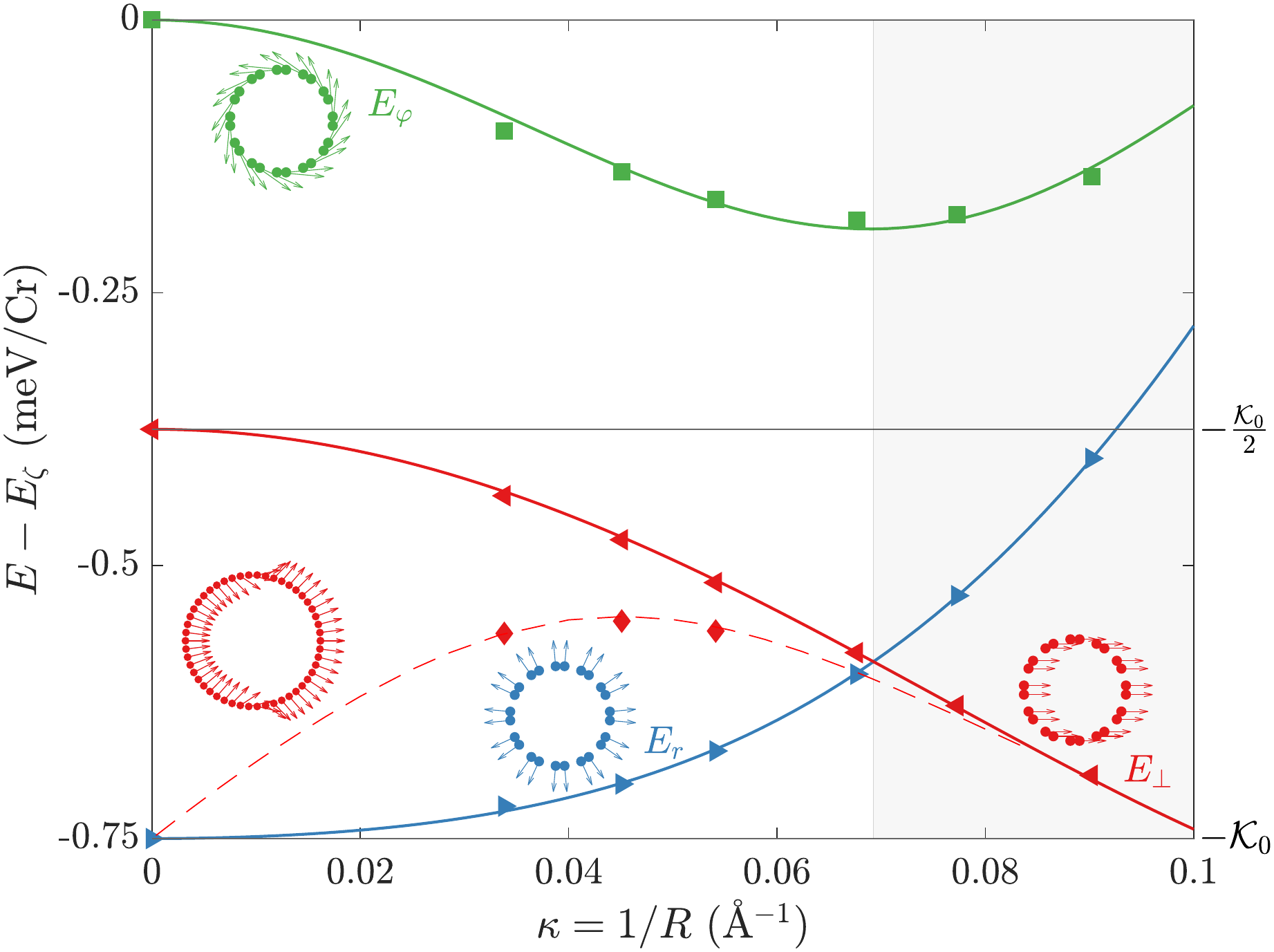}
	\caption{ Energies $E_\perp$, $E_\varphi$ and $E_\mathrm{r}$,  relative to $E_\zeta$, as functions of curvature $\kappa$, for the different states schematically represented in the insets, with magnetization along $\hat{\mathbf{e}}_\perp \perp \hat{\zeta}$, $\hat{\varphi}$ and $\hat{r}$, respectively. 
	Solid lines show energies from the continuum model discussed in the text. The dashed line shows the energy of a two domain-state~\cite{suppl} and red diamonds the corresponding DFT energies.
	}
	\label{fig.E_of_r}
\end{figure}

In Fig.~\ref{fig.E_of_r}, we show the energies of the three main spin states that we focus on in this work. 
They are magnetized either along the azimuthal direction $\hat{\varphi}$ ($E_\varphi$), the radial direction $\hat{r}$ ($E_r$), 
or with all spins parallel to a direction perpendicular to $\hat{\zeta}$ ($E_\perp$).
All energies are relative to the axial state with spins aligned along the $\hat{\zeta}$-axis of the NT. 
In the large $R$ limit we recover the known ground state of the flat monolayer, which is ferromagnetic with out-of-plane (OP) magnetization. 
This state is stabilized by an effective magnetic anisotropy constant, $\mathcal{K}_0=E_\mathrm{IP}-E_\mathrm{OP}=0.75~\mathrm{meV}$, 
defined and calculated as the energy difference between in-plane (IP) and OP magnetization in the flat monolayer.~\cite{SIAcomment} This corresponds to the energy difference between either azimuthal or axial (IP) magnetization and radial (OP) magnetization, in the small curvature limit. 
In the same limit, $E_\varphi$ tends to zero, as the azimuthal and axial one states become degenerate, while $E_r$ approaches 
$-\mathcal{K}_0$, and $E_\perp$, tends to $-\mathcal{K}_0/2$ for reasons clarified shortly.
At smaller $R$, $E_\varphi$ decreases slightly and then increases again, overall remaining higher in energy than the other two spin states.
Meanwhile, $E_r$  increases monotonously with curvature, as the angle between neighbouring spins (and hence the exchange energy) increases.
Interestingly, $E_\perp$ decreases, eventually leading to a crossover between the two spin configurations at $\kappa \approx 0.07$~\AA$^{-1}$, where $E_r = E_\perp$. 
At larger curvatures,  $E_r > E_\perp$, with the latter state becoming the lowest in energy among those considered here.  

The possibility for curvature to profoundly affect magnetism, even leading to new chiral or topological magnetic structures, 
has been pointed out before~\cite{hertel,Streubel_2016}.
The magnetic crossover shown in Fig.~\ref{fig.E_of_r} is indeed reminiscent of those discussed in Ref.~\cite{Sheka_2015}, for
opposite sign of $\mathcal{K}_0$. 
To facilitate the discussion of our results in the context of existing continuum models,
we project our magnetic configurations onto a cylindrical (locally orthogonal) coordinate system 
[see Fig.~\ref{fig.CrI3}(c)], where ($\zeta$, $\varphi$) span the tangential plane of the NT surface,
and $r$ is normal to it. 
Within the ($\zeta$, $\varphi$, $r$) system, the ``radial'' and ``azimuthal'' magnetization states correspond to constant values of either $m_r$ or $m_\varphi$,
while the ``perpendicular'' state shows a periodic out-of-phase modulation of $m_r$ and $m_\varphi$ along the
tube circumference.
More precisely, the latter state acquires the mathematical form of a spin cycloid, where
$\mathbf{m}_i = \left[ 0, -\sin(q y_i) , \cos(q y_i) \right]$ with propagation vector $q=1/R$;
an illustration is provided in Fig.~\ref{fig.CrI3}(e)-(f).
The equal mixture of azimuthal and radial spin components in the perpendicular state
explains why $E_\perp$ tends to $-K_0/2$ in the large-radius limit.
Remarkably this also means, based on the results of Fig.~\ref{fig.E_of_r}, that curvature leads to a transition
to a cycloidal magnetic ground state in the curvilinear frame of the bent surface.

\begin{figure}
	\centering
	\includegraphics[width=0.48\textwidth]{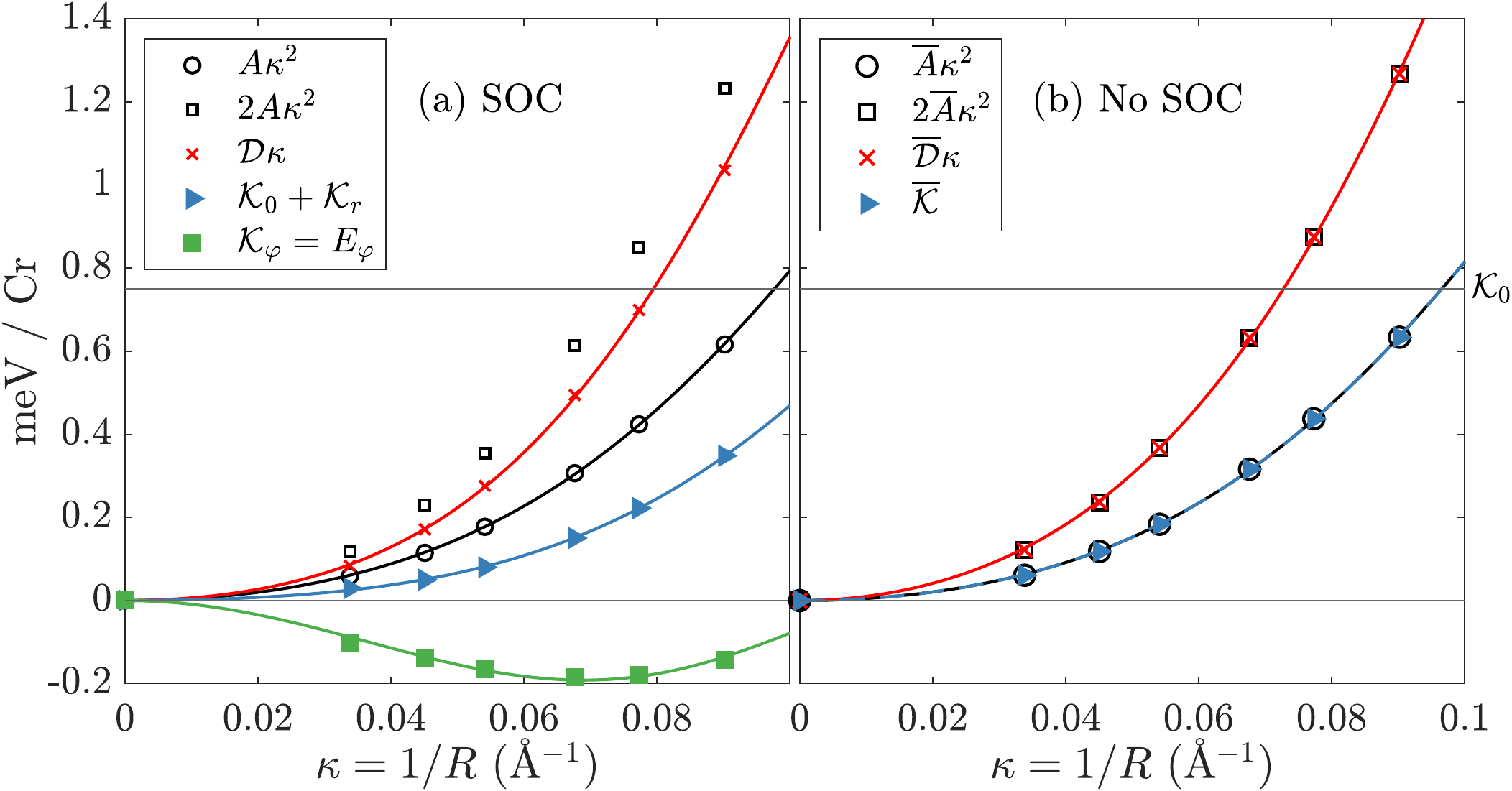}
	\caption{Energy contributions $A\kappa^2$, $\mathcal{D}\kappa$ and $\mathcal{K}$, from the effective exchange, DMI and anisotropy terms, with SOC in (a) and without SOC in (b). Lines show fits of even, 6th order polynomials. %
	}
	\label{fig.A_D_K}
\end{figure}

In order to understand the origin of such a behavior,
we consider the following continuum energy density, %
\begin{align}
    \varepsilon = &  A \left[ \partial_\varphi m_\alpha \right]^2  + \mathcal{K}_\varphi m_\varphi^2 + \mathcal{K}_r m_r^2 + 
    \nonumber \\ & +  \mathcal{D} \left[ m_r \partial_\varphi m_\varphi -m_\varphi \partial_\varphi m_r \right], 
    \label{eq.soc_contE}
\end{align}
where $A$, $\mathcal{K}_\alpha$ and $\mathcal{D}$ are the spin stiffness, anisotropy and DMI parameters, $\partial_\varphi = \frac{1}{R} \frac{\partial}{\partial \varphi}$ and $m_\alpha=m_\alpha(\varphi)$ is one of the three curvilinear components of the magnetization density.~\cite{Mzeta} %
Note that all parameters in Eq.~(\ref{eq.soc_contE}) depend on curvature, e.g. $A=A(\kappa=\frac{1}{R})$. 
To extract this dependence from our DFT calculations we fit Eq.~(\ref{eq.soc_contE}), at each $R$, to the calculated energies of the
three magnetic states in Fig.~\ref{fig.E_of_r}.
The azimuthal and radial anisotropy constants are trivially 
provided by two of the three sets of data: 
$\mathcal{K}_\varphi=E_\varphi$ and $ \mathcal{K}_{r} = E_r $.
Separating the remaining two parameters ($A$ and $\mathcal{D}$) is computationally more involved,
since their contributions to the energy of the perpendicular state, 
$E_\perp = A\kappa^2 - \mathcal{D}\kappa  + \frac{1}{2}(\mathcal{K}_r + \mathcal{K}_\varphi)$,
are linearly dependent.
(By symmetry, $A$ and $\mathcal{D} \kappa$ are both even functions of $\kappa$.)
To extract also $A(\kappa)$ 
and $\mathcal{D}(\kappa)$, we additionally perform calculations for
cycloidal states of the form $\mathbf{m} = \cos(n\varphi)\hat{r} - \sin(n\varphi)\hat{\varphi}$, 
with integer values $n>1$, consistent with $2\pi$-periodicity in $\varphi$. 
($n=0$ and $n=1$ correspond to the radial and perpendicular magnetization states, already described.)
This procedure allows us to uniquely resolve $A(\kappa)$, $\mathcal{D}(\kappa)$ and  $\mathcal{K}(\kappa)$
at the discrete set of curvatures considered in our DFT calculations.
For practical purposes, we then interpolate these data with appropriate low-order polynomials 
of $\kappa$ (See Fig.~\ref{fig.A_D_K} and the SI~\cite{suppl}), which yields the continuous curves plotted in Fig.~\ref{fig.E_of_r}.

The resulting separate energy contributions $A\kappa^2$, $\mathcal{D}\kappa$ and $\mathcal{K}_\alpha$ from the effective spin stiffness, DMI and anisotropy parameters are plotted as functions of curvature in Fig.~\ref{fig.A_D_K}(a). 
This decomposition allows us to clarify 
the physical origin of the cycloidal ground state obtained at larger $\kappa$.
The increase in the spin stiffness energy with curvature leads to an increase in the energy cost of the spin cycloids, relative to a FM state. A non-zero $\mathcal{D}$ (linear in $\kappa$), however, also develops with curvature, favoring the crossover to the $n = 1$ cycloidal state. Both the curvature-induced $\mathcal{D}$ and the flat-layer anisotropy $\mathcal{K}_0$ are essential to stabilizing states with $n>0$ \cite{suppl}; the latter only occurs with SOC. 
Note that the largest contribution to the energy comes from the DMI term. 
At $\kappa\approx 0.1 ~\mathrm{\AA}^{-1}$, $\mathcal{D}\kappa \approx 1~\mathrm{meV/Cr}$. Thus, the value of $\mathcal{D}$ reaches almost 1~meV/\AA, similar as the interfacial DMI of a surface of Fe on W~\cite{HEIDE20092678,area}.
 For the flat, centrosymmetric CrI$_3$ monolayer, the DMI arises as an interaction between second neighbours, estimated to $60~\mathrm{\mu eV}$ or even smaller~\cite{PhysRevMaterials.4.094004,PhysRevB.102.115162}, whereas electric field ($\mathcal{E}$) induced DMI between nearest Cr-neighbours was reported to be $14~\mathrm{\mu eV}$ for $\mathcal{E}_z\simeq1.55 ~\mathrm{V/nm}$~\cite{PhysRevMaterials.4.094004}; both values are orders of magnitude smaller than the effective, curvature-induced DMI energy found here. 

With the parameters of Eq.~(\ref{eq.soc_contE}) at hand, we can calculate the stationary states, minimizing the continuum energy functional~\cite{suppl}.  
This confirms the transition from the radial to the "perpendicular" magnetization state at large curvatures. However, at small curvatures a qualitatively different behavior is seen; the "perpendicular" (n = 1 cycloid) state evolves into two domains with opposite radial magnetizations and the energy goes to $-\mathcal{K}_0$ linearly, instead of $-\mathcal{K}_0/2$ quadratically, as shown by the red dashed line in Fig.~\ref{fig.E_of_r}. For the largest $R$ NTs, we performed additional DFT calculations for the predicted spin configurations (red diamons in Fig.~\ref{fig.E_of_r}), confirming the predictions of the model with excellent accuracy.   

Earlier phenomenological studies~\cite{PhysRevLett.112.257203,PhysRevLett.114.197204,Sheka_2015,Streubel_2016} 
showed the appearance of effective anisotropy and DMI terms in curvilinear magnets, consistent with our findings. 
Remarkably, such effects were predicted even in absence of SOC.
%
To check whether such assumptions are reliable in our case, in the following 
we benchmark the model of Ref.~\cite{Sheka_2015}, which we have adapted to our specific geometry~\cite{suppl}, against our first-principles results.
Consistent with the conclusions of Sheka {\em et al.}~\cite{Sheka_2015}, we find that the coefficients of Eq.~(\ref{eq.soc_contE}) 
are no longer independent in absence of SOC, but enjoy the following mutual relationships~\cite{suppl},
\begin{equation}\label{eq.noSOCpars}
\overline{\mathcal{D}}=2\overline{A}\kappa , \qquad \overline{\mathcal{K}}=\overline{\mathcal{K}}_r=\overline{\mathcal{K}}_\varphi=\overline{A} \kappa^2.
\end{equation}
(overline symbols indicate quantities defined and 
calculated without SOC.) 
Our results of Fig.~\ref{fig.A_D_K}(a) manifestly violate such conditions: neither the effective anisotropies nor the DMI appear to be related to the spin stiffness energy 
in any obvious way.
To clarify the role of SOC in this unexpected behavior, 
we have repeated all our first-principles calculations without SOC, 
and followed the same post-processing procedure to extract 
$\overline{A}$, $\overline{\mathcal{D}}$ and $\overline{\mathcal{K}}$ as functions of curvature; the energies $\overline{A}\kappa^2$, $\overline{\mathcal{D}}\kappa$ and $\overline{\mathcal{K}}$  are shown in Fig.~\ref{fig.A_D_K}(b).
Without SOC, we find that Eq.~(\ref{eq.noSOCpars}) exactly holds within numerical accuracy. This is not unexpected as the energy no longer depends on the global quantization axis, but only on the relative angle between neighboring spins. 
This implies an exact degeneracy ($E_r=E_\varphi=E_{n=2}=\overline{A}\kappa^2$) between the ``radial'', ``azimuthal'', 
and $n=2$ cycloidal states on one hand, and between the ``axial'' and ``perpendicular'' states ($E_\zeta=E_\perp=0$) 
on the other. These combined facts prove that
the aforementioned violation [Fig.~\ref{fig.A_D_K}(a)] of Eq.~(\ref{eq.noSOCpars})  
is entirely due to SOC. From Eq.~(\ref{eq.soc_contE}), one finds~\cite{suppl} that $\varepsilon_{n}<\varepsilon_{n-1}$ if $\mathcal{D}>(2n-1)A\kappa$. Specifically, $\varepsilon_{n=2}<\varepsilon_{n=1}$ if $\mathcal{D} > 3A\kappa$. This can be compared to the above relation $\overline{\mathcal{D}} = 2\overline{A}\kappa$, whereby a SOC enhancement of $\mathcal{D}$ by 50\%, relative to $A$, could stabilize the $n=2$ state, in materials with strong SOC.

\begin{table*}[t]
\caption{Fitted leading order coefficients (energy per Cr) for the curvature dependence of spin stiffness, anisotropy and DMI.~\cite{Aspiral} }
\label{table}
\begin{tabular}{lccccccc}
\hline \hline
         & $A_0$ (meV\AA$^{2}$) & [$A_0$ (meV\AA$^{2}$)] & $A_2$  (eV\AA$^{4}$) & $\mathcal{K}_0$ (meV) & $\mathcal{K}_{r,2}$ (meV\AA$^{2}$) & $\mathcal{K}_{\varphi,2}$ (meV\AA$^{2}$) & $\mathcal{D}_1$ (meV\AA$^2$) \\ \hline
No SOC   & 49.9   &  42.9 & 4.9  & 0        & 49.9      & 49.9     & 99.9   \\
With SOC &  48.1  & 40.7 & 4.8   & 0.75   & 17.7      & -90.4    &   63.4 \\
\hline \hline
\end{tabular}
\end{table*}

The above analysis reveals an impact of relativistic effects on the curvilinear spin Hamiltonian, far more profound than previously believed.
Existing models~\cite{Sheka_2015,PhysRevLett.112.257203,PhysRevLett.114.197204} limit their account of relativistic effects 
to including a curvature-independent anisotropy constant $K_0$, while any impact of SOC on effective anisotropy and DMI 
interactions is systematically neglected.
Our results show that the DMI is reduced in magnitude by around 30\% by SOC, while
the two anisotropy constants $\mathcal{K}_{\varphi }$ and $\mathcal{K}_{r}$ display a very different curvature 
dependence in presence of SOC, in stark disagreement with the non-relativistic relationships of Eq.~(\ref{eq.noSOCpars}). 
Remarkably, our calculated $\mathcal{K}_\varphi$ is \emph{opposite in sign}
to what would result from the physics of isotropic exchange alone.
This shows that SOC needs to be taken into account in the calculation of the effective anisotropy of 
bent layers, as its neglect might lead to qualitatively wrong physical answers.
For a quantitative comparison, in Table~\ref{table} we list the lowest-order 
fitted coefficients of $A(\kappa)$, $\mathcal{K}_\alpha (\kappa)$ and $\mathcal{D}(\kappa)$, 
with and without SOC.
Clearly, the spin stiffness ($A_0$ and $A_2$) is not substantially affected by SOC. 
The effective DMI,  
$\overline{\mathcal{D}}_1 = 2\overline{A}_0 = 99.9~\mathrm{meV\AA}^{2}\mathrm{/Cr}$ is increased compared to $\mathcal{D}_1 = 63.4~\mathrm{meV\AA}^{2}\mathrm{/Cr}$ with SOC. 
The leading order change in $K_\varphi$, would correspond to $\overline{A}_0 = 49.9$ meV\AA$^{2}$/ Cr in the view of earlier models, in clear contrast to the value of $-90.4$ meV\AA$^{2}$/ Cr found from our DFT calculations including SOC.

\emph{Conclusions - } 
We have used non-collinear magnetic DFT calculations, with and without SOC, to investigate the interplay of curvature and magnetism in monolayer CrI$_3$.
In addition to a crossover between two spin states of distinct symmetry, our calculations demonstrated that it is essential to take into account the effects of SOC for a quantitatively (and sometimes even qualitatively) accurate description of the flexomagnetic coupling parameters. 
The obvious question is whether these conclusions are specific to the material considered here, or
whether they are relevant to a broader range of systems. We can't give a definite answer at
this stage, but we can certainly speculate on how effective the present strategy may be in studying 
other cases.
The main limitation we see in this context is related to computational power:
While adapting our method to other 2D layers appears straightforward, the study of thicker membranes
may be out of reach at present, due to the costly NT geometry. 
A way forward may be provided by the so-called cyclic DFT method,~\cite{BANERJEE2016605} 
which allows for calculating bent structures 
at a significantly lower computational cost; whether such an approach is 
effective in non-collinear spin structures, however, remains to be seen.
An alternative possibility would be treating curvature perturbatively via flexural phonons, in analogy
to the ongoing efforts in the theory of flexoelectricity~\cite{Stengel_Vanderbilt}; this could possibly allow one to work with
the primitive cell of the flat crystal, with considerable savings in computer power.
Exploration of these promising avenues, together with the discussion of 
other related physical effects going beyond the Hamiltonian of Eq.~(\ref{eq.soc_contE}), 
will be an exciting topic for future studies.

\begin{acknowledgments}
\emph{Acknowledgments - } We acknowledge financial support from the Swedish Research Council (VR - 2018-06807).  M.S. acknowledges the support of Ministerio de Economia,
Industria y Competitividad (MINECO-Spain) through
Grants No. PID2019-108573GB-C22 and Severo Ochoa FUNFUTURE 
center of excellence (CEX2019-000917-S);
of Generalitat de Catalunya (Grant No. 2017 SGR1506);
and of the European Research Council (ERC) under the European Union's
Horizon 2020 research and innovation program (Grant
Agreement No. 724529).
P.B. and S.P. acknowledge financial support from Italian MIUR under the PRIN project "Tuning and understanding Quantum phases in 2D materials - Quantum2D", grant n.  2017Z8TS5B and "TWEET: Towards Ferroelectricity
in two dimensions", grant n. 2017YCTB59, respectively.  D.A. and S.P. acknowledge financial support by the Nanoscience Foundries and Fine Analysis (NFFA-MIUR Italy) project.
The authors thankfully acknowledge the computer resources at Pirineus and the technical support provided by CSUC (RES-FI-2021-1-0034). Additionally, computational work was done on resources at PDC, Stockholm via the Swedish National Infrastructure for Computing (SNIC).
\end{acknowledgments}

\bibliography{literature}{}

\begin{thebibliography}{58}
\bibitem [{Mze()}]{Mzeta}%
  \BibitemOpen
  \href@noop {} {}\bibinfo {note} {We consider $m_\alpha$ as a constant
  function of the axial $\zeta$ coordinate.}\BibitemShut {Stop}%
\bibitem [{Asp()}]{Aspiral}%
  \BibitemOpen
  \href@noop {} {}\bibinfo {note} {The value of $A_0$ in brackets was obtained
  from spin cycloid energies of the flat monolayer, showing a moderate
  numerical discrepancy to the value fitted including the curved
  systems.}\BibitemShut {Stop}%
\end{thebibliography}
\bibliographystyle{apsrev4-1}

\end{document}


\title{Supplemental Material: Curved Magnetism in CrI$_3$}
 
\author{Alexander Edstr\"om}
\affiliation{Institut  de  Ci\`encia  de  Materials  de  Barcelona  (ICMAB-CSIC),  Campus  UAB,  08193  Bellaterra, Spain}
\affiliation{Department of Applied Physics, School of Engineering Sciences,KTH Royal Institute of Technology, AlbaNova University Center, 10691 Stockholm, Sweden}
\author{Danila Amoroso}
\affiliation{Consiglio Nazionale delle Ricerche CNR-SPIN, c/o Universit\`{a} degli Studi 'G. D'Annunzio', 66100, Chieti, Italy}
\affiliation{NanoMat/Q-mat/CESAM, Universit\'e de Li\`ege, B-4000 Liege, Belgium}
\author{Silvia Picozzi}
\affiliation{Consiglio Nazionale delle Ricerche CNR-SPIN, c/o Universit\`{a} degli Studi 'G. D'Annunzio', 66100, Chieti, Italy}
\author{Paolo Barone}
\affiliation{Consiglio Nazionale delle Ricerche CNR-SPIN,  Area della Ricerca di Tor Vergata,Via del Fosso del Cavaliere 100, I-00133 Rome, Italy}
\author{Massimiliano Stengel}
\affiliation{Institut  de  Ci\`encia  de  Materials  de  Barcelona  (ICMAB-CSIC),  Campus  UAB,  08193  Bellaterra, Spain}
\affiliation{ICREA  -  Instituci\`{o}  Catalana  de  Recerca  i  Estudis  Avan\c{c}ats,  08010  Barcelona,  Spain}

\begin{abstract}
This supplemental material contains a detailed description of the fitting procedure used to obtain the parameters of the magnetic continuum model from the DFT data. Additionally, calculations without constraints on the spin directions, comparison to other continuum models, and a discussion of stationary states of the continuum energy, and the related formation of magnetic domains, are discussed.
\end{abstract}

\maketitle

\subsection{Fitting $A$, $\mathcal{D}$ and $\mathcal{K}$} 

As introduced in the main text, we consider a continuum magnetic energy density 
\begin{equation}
    \varepsilon = A_{\zeta} \left[ \partial_\zeta m_\alpha \right]^2 + A_\varphi \left[ \partial_\varphi m_\alpha\right]^2  + \mathcal{K}_\varphi m_\varphi^2 + \mathcal{K}_r m_r^2 + \mathcal{K}_\zeta m_\zeta^2 +  \mathcal{D} \left[ m_r \partial_\varphi m_\varphi -m_\varphi \partial_\varphi m_r \right]. 
    \label{eq.soc_cont_mod}
\end{equation}
Additional DM terms are allowed, but not relevant for the magnetic states considered here. We define $\mathcal{K}_{\zeta}=0$. Further, we only consider states where $\partial_\zeta \mathbf{m}=0$, whereby the first term vanishes and we call $A_\varphi=A$. $\alpha$ runs over the curvilinear coordinates and is summed over. The total magnetic energy is 
\begin{equation} \label{eq.totE}
    E = \int_S \varepsilon [ \mathbf{m}(\varphi,\zeta), \partial_\alpha \mathbf{m}(\varphi,\zeta)] \dd S, 
\end{equation}
with integration over the surface $S$. 

\begin{figure}
	\centering
	\includegraphics[width=.6\textwidth]{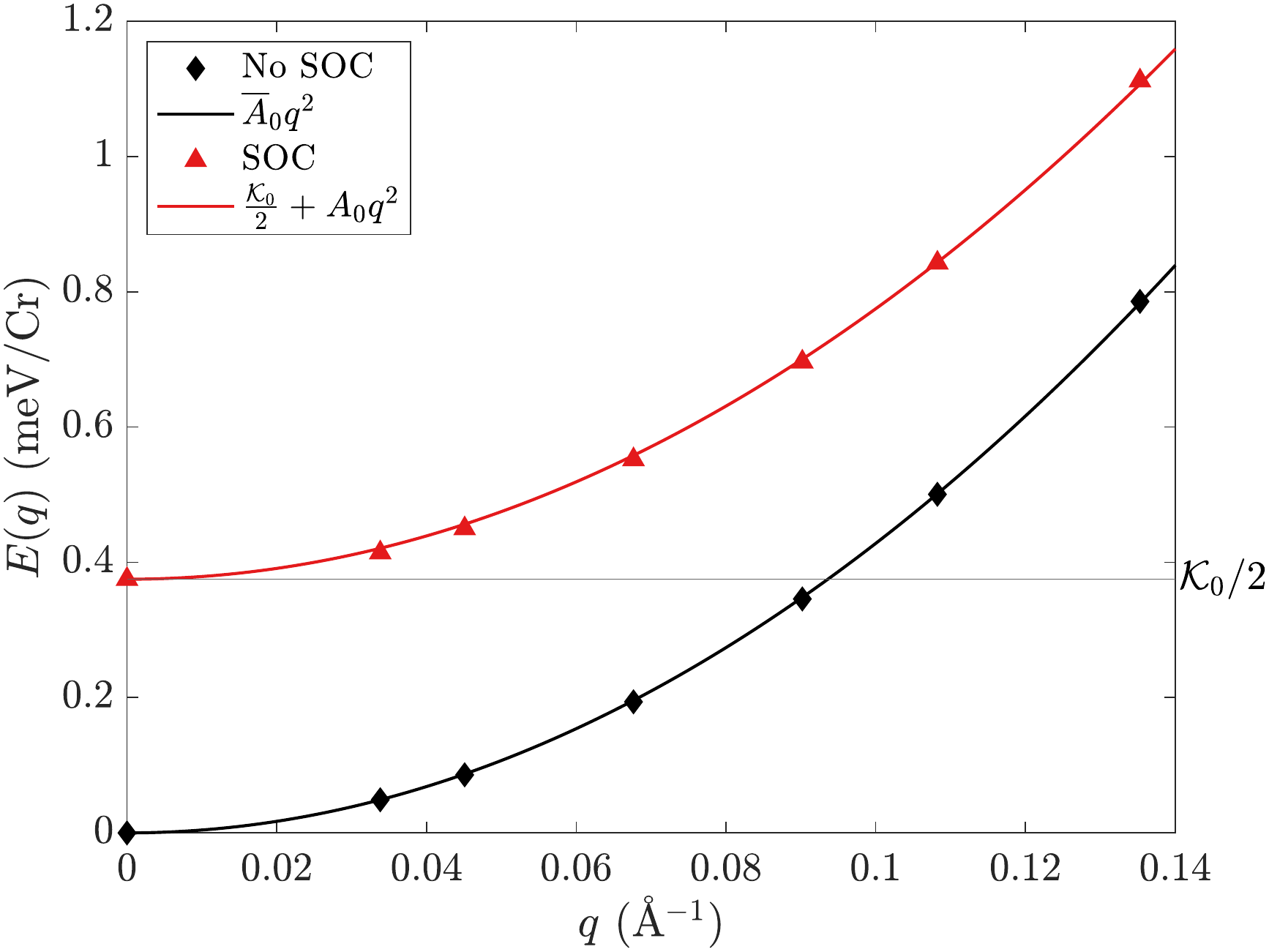}
	\caption{Energy as function of propagation vector $q$ for cycloidal spin spirals of the form $\mathbf{m}_i = \left[ 0, \sin(q y_i) , \cos(q y_i) \right]$, in a flat monolayer.  }
	\label{fig.flat_spirals}
\end{figure}

The value $A_0$, of $A$ in the zero curvature limit, can be determined from spin spiral calculations in a flat monolayer. The total energy as function of the propagation vector $q$ is shown, with and without spin-orbit coupling (SOC), in Fig.~\ref{fig.flat_spirals}, together with fits of the form $E=-\frac{\mathcal{K}_0}{2} + A_0 q^2$ with SOC and $E= \overline{A}_0 q^2$ without SOC, leading to the values $A_0 = 40.7~\mathrm{meV\AA^{2}/ Cr}$ and $\overline{A}_0 = 42.9~\mathrm{meV\AA^{2}/ Cr}$.

\begin{figure}
	\centering
	\includegraphics[width=1.0\textwidth]{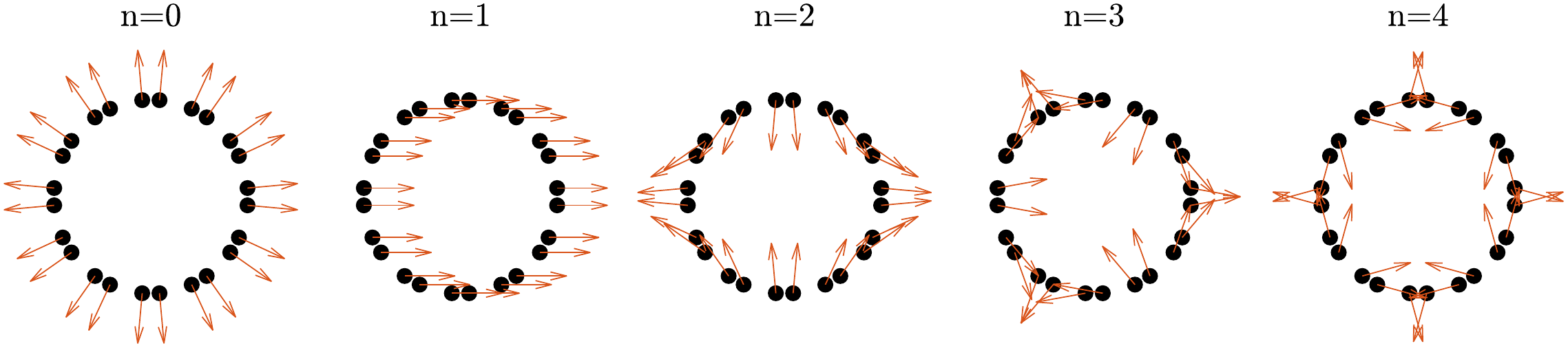}
	\caption{Illustration of the nanotube cycloidal spin spiral states characterized by $n=qR$, for the $N=6$ nanotube. Dots indicate Cr atoms and arrows their spins. Note that $n=0$ corresponds to the radial magnetization state, and $n=1$ to the state with spins parallel to each other but perpendicular to the tube axis.}
	\label{fig.tube_spirals}
\end{figure}

Considering a nanotube with circumference $L=2\pi R$, and the radial magnetization state being analogous to the out-of-plane ferromagnetic state (the ground state) of a flat monolayer, then a magnetization $\mathbf{m} = \cos(n\varphi)\hat{r} - \sin(n\varphi)\hat{\varphi}$ corresponds to a cycloidal spin spiral state with wavevector $q=\frac{2\pi n}{L}=\frac{n}{R}=n\kappa$. We calculated DFT total energies of these states for $0 \leq n \leq 4$, as illustrated in Fig.~\ref{fig.tube_spirals}. For non-zero, integer values of $n$, the energy is 
\begin{align}
    \varepsilon_n = A\kappa^2n^2 -\mathcal{D}\kappa n + \frac{1}{2}(\mathcal{K}_\varphi + \mathcal{K}_r).
    \label{Eq.E_n}
\end{align}
For $n=0$, which is the radial magnetization state, with SOC there is an extra contribution $(\mathcal{K}_{r}-\mathcal{K}_{\varphi })/2$, so the energy is $\varepsilon_0 = \mathcal{K}_{r}$. Note that the linear dependence on $n$ from the effective DMI term is a magnetochiral effect, whereby the energy depends on the direction of rotation of the spins. Fitting DFT calculated energies as function of $n$ allows $A$, $\mathcal{D}$, $\mathcal{K}_{r}$ and $\mathcal{K}_{\varphi}$ to be evaluated. However, $\mathcal{K}_{r}$ and $\mathcal{K}_{\varphi}$ are already known from the data in Fig.~2 in the main text, leaving only $A$ and $\mathcal{D}$ to be determined. In the case with SOC, the energy of the axial magnetization state is additionally needed to define the zero energy. %

\begin{figure}
	\centering
	\includegraphics[width=0.88\textwidth]{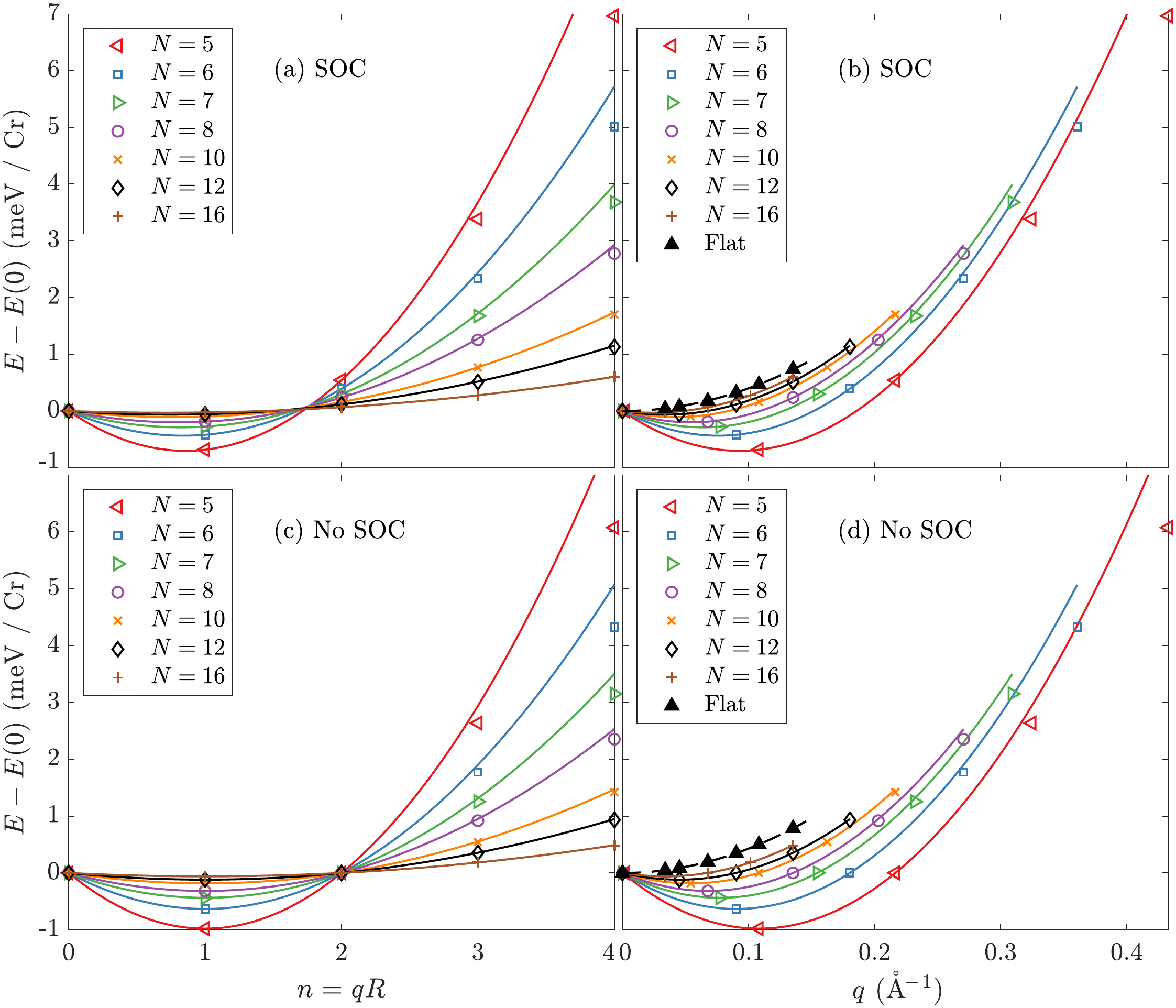}
	\caption{Energy as function of $n=qR$ in (a) and (c) and of $q$ in (b) and (d), for different nanotube radii $R=NR_0$, also showing quadratic fits for $0 \leq n \leq 2$ with solid lines. Data with SOC is shown in (a)-(b), and without SOC in (c)-(d). The data with SOC have been shifted by $E(0)=-\mathcal{K}_{r}/2$, known from Fig.~2 in the main text. }
	\label{fig.E_of_n_etc}
\end{figure}

The fittings of $E(n,N)=E(q,R)$ used to determine $A$ and $\mathcal{D}$ are plotted in Fig.~\ref{fig.E_of_n_etc}. For small radii ($N$) and large wavevectors ($q \propto n$) the continuum approximation loses its validity. Hence, the fitting is done only using $n\leq 2$. For large $N$ the fitted curves still agree well with the data points at $n=3$, while there is an increasing discrepancy for decreasing $N$, as the continuum approximations breaks down. The same trend is seen at $n=4$, but with an increased discrepancy between fitted curves and calculated points, compared to $n=3$. 

In the case without SOC, the spin states corresponding to $n=0$ and $n=2$ are degenerate. This is because these states have the same magnitude of the angles between neighboring spins. However, including SOC breaks this degeneracy by introducing magnetic anisotropy.  

\begin{figure}
	\centering
	\includegraphics[width=0.9\textwidth]{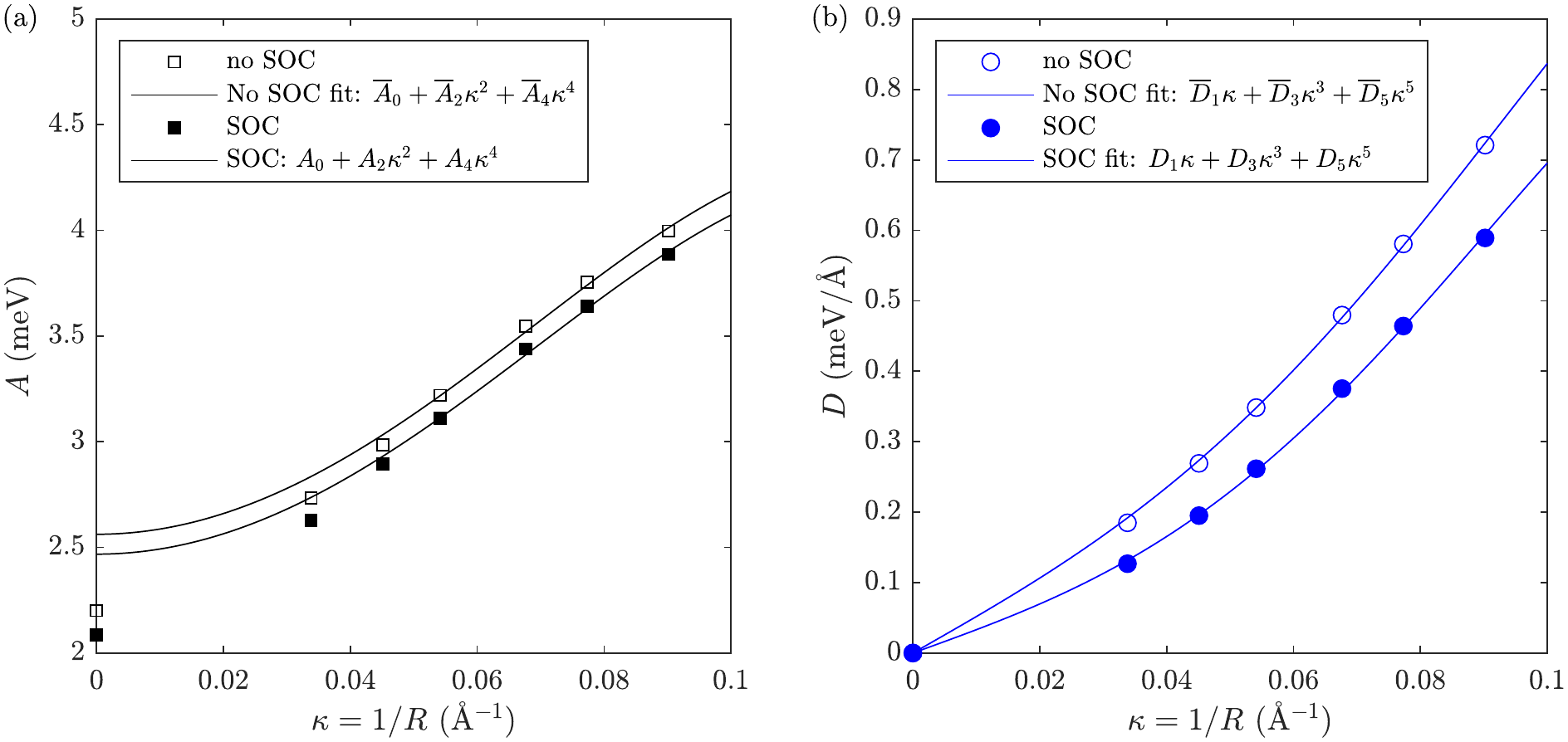}	
	\caption{(a) Spin stiffness $A(\kappa)$ and (b) DMI $\mathcal{D}(\kappa)$ as functions of curvature, with and without SOC. 
	}
	\label{fig.compare_A_K_D}
\end{figure}

The extracted values for the spin stiffness $A$ and DMI parameter $\mathcal{D}$, with and without SOC, are plotted as functions of curvature in Fig.~\ref{fig.compare_A_K_D}. The values extracted for the anisotropy constants are plotted in Fig. 2 and Fig. 3 of the main text. By symmetry, the spin stiffness and anisotropy are even functions of curvature, while the effective DMI is odd. Consequently, they all contribute to the total energy as even functions of curvature (DMI contributes to the energy as $\mathcal{D}\kappa$). As discussed in the main text, and seen in Fig.~\ref{fig.compare_A_K_D}, $A$ is barely affected by SOC. Meanwhile, also the DMI behaves qualitatively similarly with or without SOC, although SOC reduces the magnitude of $\mathcal{D}$ compared to the case without SOC. 

{
The curvature dependence of the parameters is interpolated using a polynomial fitting
\begin{align}
A(\kappa)\kappa^2 & = A_0 \kappa^2 + A_2 \kappa^4 + A_4 \kappa^6 \label{eq.Afit} \\
\mathcal{D}(\kappa)\kappa & = \mathcal{D}_1 \kappa^2 + \mathcal{D}_3 \kappa^4 + \mathcal{D}_5 \kappa^6  \\
\mathcal{K}(\kappa) - \mathcal{K}(0) & = \mathcal{K}_2 \kappa^2 + \mathcal{K}_4 \kappa^4 + \mathcal{K}_6 \kappa^6 , \label{eq.Kfit}
\end{align}
which treats the parameters at equal footing and preserves the expected relations between the coefficients in the case without SOC. The fittings of Eqs.~(\ref{eq.Afit})-(\ref{eq.Kfit}) are plotted with the solid lines in Fig. 3 of the main text, where it is seen that the corresponding energies are excellently described by the fit. As seen in Fig.~\ref{fig.compare_A_K_D}, the fit also gives an excellent description of $\mathcal{D}$, over the curvatures considered. In the case of $A$, the agreement of the fit is excellent for larger curvatures, while a numerical discrepancy is seen for small curvatures, resulting in a difference between the $A_0 = 49.9~\mathrm{meV\AA^{2} / Cr}$ and $\overline{A}_0 = 48.1~\mathrm{meV\AA^{2} / Cr}$ obtained from this fit and the values of $A_0 = 42.9~\mathrm{meV\AA^{2} / Cr}$ and $\overline{A}_0 = 40.7~\mathrm{meV\AA^{2} / Cr}$ from the fit of the data in Fig.~\ref{fig.flat_spirals}. However, as confirmed in Fig. 3 of the main text, this moderate numerical discrepancy in $A$ does not affect the relevant energy to any significant extent.  
}

Based on Eq.~(\ref{eq.soc_cont_mod}), the relevant energies which have been calculated with DFT, considered in the main text, are $\varepsilon_r = \varepsilon_{n=0} = \mathcal{K}_r$, $\varepsilon_\varphi = \mathcal{K}_\varphi$, $\varepsilon_\zeta = \mathcal{K}_\zeta = 0$ and 
\begin{equation}
    \varepsilon_\perp = \varepsilon_{n=1} = \kappa^2 A - \kappa\mathcal{D} + \frac{1}{2}(\mathcal{K}_\varphi + \mathcal{K}_r).
\end{equation}
Using the $\kappa$-dependence of $A$ and $\mathcal{D}$ shown in Fig~\ref{fig.compare_A_K_D}, and the fittings for $\mathcal{K}_r$ and $\mathcal{K}_\varphi$, the solid lines in Fig.~2 of the main text are drawn.

{ From Eq.~(\ref{Eq.E_n}) one can see that non-zero, positive values of $n$ are favored by the effective DMI term. Large values of $\mathcal{D}$ could therefore also stabilize states with $n>1$. A state with $n>1$ is lower in energy than that with $n-1$ if the condition $\mathcal{D}>(2n-1)A\kappa$ is fulfilled. In particular, the $n=2$ state is lower in energy than the $n=1$ state if $\mathcal{D} > 3A\kappa$;  the latter condition should be compared to the relation $\overline{\mathcal{D}} = 2\overline{A}\kappa$ fulfilled without SOC (see derivation below). Hence, the $n=2$ state could be stabilized in materials where SOC increases the value of $\mathcal{D}$ relative to $A$ by 50\%, compared to the case without SOC. In principle, higher values of $n$ could also be stabilized by a large $\mathcal{D}$, although in practice it may require an unrealistically strong effect of SOC. }

Without SOC, the energy is determined by the isotropic exchange term, which in the continuum/long wavelength limit is $\varepsilon_\mathrm{ex} = \overline{A} \left[ \nabla \mathbf{m} \right]^2$. In the curvilinear coordinates $\zeta, \varphi, r $, for a cylinder with radius $R$, 
\begin{align} \label{eq.exch}
    \varepsilon_\mathrm{ex} & = \overline{A} \left[ \partial_\varphi m_\alpha \right]^2 + {\overline{A}{R^2} \left[ m_r^2 +  m_\varphi^2  \right]} +  \frac{2\overline{A}}{R} \left[ m_r \partial_\varphi m_\varphi -m_\varphi \partial_\varphi m_r  \right] \\ 
    & = \varepsilon_\mathrm{ex}^\mathrm{eff} + \underbrace{\overline{\mathcal{K}} \left[ m_r^2 +  m_\varphi^2  \right]}_{ \varepsilon_\mathrm{ani}^\mathrm{eff}} + \underbrace{ \overline{\mathcal{D}} \left[ m_r \partial_\varphi m_\varphi -m_\varphi \partial_\varphi m_r \right] }_{ \varepsilon_\mathrm{DMI}^\mathrm{eff}},
\end{align}
with effective exchange, anisotropy and Dzyaloshinskii–Moriya interaction (DMI) terms. Accordingly, the curvature induced, non-relativistic, anisotropy and DMI constants are $\overline{\mathcal{K}}=\frac{\overline{A}}{R^2}$ and $\overline{\mathcal{D}}=\frac{2\overline{A}}{R}$, respectively. Note that $\overline{A}=\overline{A}(\kappa=1/R)$ is also curvature dependent. As seen in Fig.~3 of the main text, the relations $\overline{\mathcal{K}}=\frac{\overline{A}}{R^2}$ and $\overline{\mathcal{D}}=\frac{2\overline{A}}{R}$ are excellently  fulfilled by the fitted values for $\overline{A}$, $\overline{\mathcal{D}}$ and $\overline{\mathcal{K}}$. %

{
\subsection{Stationary States of the Continuum Energy Functional}

Having determined the curvature dependent parameters of the continuum energy functional, we can proceed to find the stationary states minimizing the energy. Considering a magnetization $\mathbf{m} = \cos(\theta)\hat{z} + \sin(\theta)\hat{y}$, where $\theta(y)$ is the angle between the magnetization and the normal direction at $y$, the energy density of Eq.~(\ref{eq.soc_cont_mod}) takes the form 
\begin{align}
    \varepsilon_\mathrm{flat} & = A_0 \left(\frac{\dd \theta}{\dd y}\right)^2 - \mathcal{K}_0 \cos^2 \theta  \label{eq.soc_contE_th} \\
    \varepsilon_\mathrm{curved} & = \frac{A}{R^2} \left(\frac{\dd \theta}{\dd \varphi}\right)^2 + \mathcal{K}_\varphi \sin^2 \theta + \mathcal{K}_r \cos^2 \theta + \frac{\mathcal{D}}{R} \frac{\dd \theta}{\dd \varphi}
    \label{eq.soc_contE_th_curve}.
\end{align}
with the first line written in the Cartesian coordinates appropriate for the flat monolayer, and the second line written in the curvilinear coordinates appropriate for the curved system, related via $y=R\varphi$. 
Minimization of the energy [Eq.~(\ref{eq.totE})] using the Euler-Lagrange equation, leads to the stationary Sine-Gordon equation 
\begin{align}
\frac{\dd^2 \theta}{\dd y^2}  & = \frac{\mathcal{K}_0}{2A_0}  \sin 2\theta \label{eq.sineGord},  \\
\frac{\dd^2 \theta}{\dd \varphi^2}  & = \frac{(\mathcal{K}_\varphi - \mathcal{K}_r)R^2}{2A}  \sin 2\theta, 
\label{eq.sineGord_curv} 
\end{align}
for the flat and curved cases respectively. 

In the flat case, applying the boundary conditions $\theta(-\infty) = 0 $ and $\theta(\infty) = \pi $, leads to a domain wall (DW) solution of Eq.~(\ref{eq.sineGord})
\begin{equation}\label{eq.flatDW}
    \theta(y) = 2 \arctan(\ee^{y/\xi}) ,
\end{equation}
with DW width 
\begin{equation}
    \xi = \sqrt{\frac{A_0}{\mathcal{K}_0}} = 8~\mathrm{Å}.
\end{equation}
Integration of Eq.~(\ref{eq.soc_contE_th}), using the DW solution in Eq.~(\ref{eq.flatDW}), leads to a DW energy of 22.0 meV/\AA.

For the curved case with nanotube geometry, the boundary conditions $\theta(0) = 0$ and $\theta(2\pi) = -2n\pi$ correspond to those of the $q=\frac{n}{R} = n\kappa$ spin cycloids discussed above. However, as will be seen below, for large radii $R$ (small curvatures $\kappa$) the corresponding solutions of Eq.~(\ref{eq.sineGord_curv}) evolve into $2n$-domain states, with opposite domains having $\mathbf{m} = \hat{r}$ and $\mathbf{m} = -\hat{r}$.   

We solved Eq.~(\ref{eq.sineGord_curv}) numerically, using a finite difference approach, for different curvatures/radii, using the curvature dependent values of the parameters ($A$, $\mathcal{K}_r$, $\mathcal{K}_\varphi$,$\mathcal{D}$), presented in Fig.~\ref{fig.compare_A_K_D} and in the main text. From the solution for $\theta(\varphi)$, the corresponding energy is calculated by integrating Eq.~(\ref{eq.soc_contE_th_curve}). This approach allows us to extract information about the behavior at small curvatures, beyond the computational capabilities of the DFT calculations.

Fig.~\ref{fig.E_of_curv_numsol} shows the energy as a function of curvature calculated by minimizing the energy of the continuum energy functional for the periodic boundary conditions with different values of $n$. For small curvatures, the lowest energy solution is that with $n=0$, which at every curvature considered keeps the magnetization exactly along the radial direction. At larger curvature, there is a transition and the lowest energy state becomes that with $n=1$, in agreement with the DFT data. 

\begin{figure}
	\centering
	\includegraphics[width=0.72\textwidth]{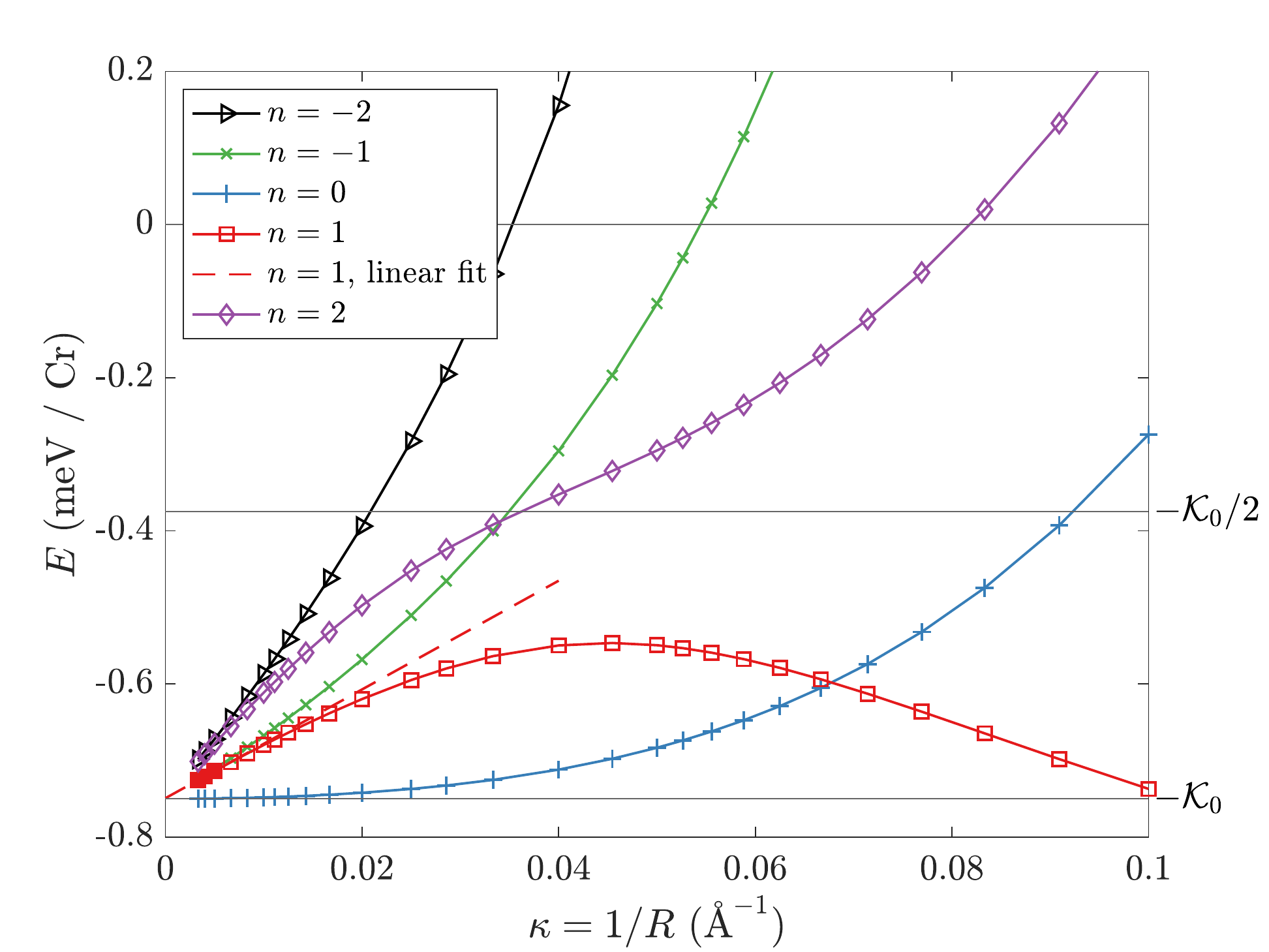}
	\caption{Energy as a function of curvature, calculated by numerically minimizing the continuum energy functional for boundary conditions characterized by $n$. The dashed line shows a linear fit at small curvatures for $n=1$.  }
	\label{fig.E_of_curv_numsol}
\end{figure}

However, going towards small curvatures, corresponding to radii beyond that of $R=30~\mathrm{\AA}$ ($\kappa=0.033~\mathrm{\AA^{-1}}$) which was studied with DFT, one can see qualitatively new behaviour in the states with $n \neq 0$; Instead of going towards $-\mathcal{K}_0/2$ quadratically as the curvature approaches zero, the energies go towards $-\mathcal{K}_0$ linearly for small curvatures. This is because, when the diameter is significantly larger than $2n\xi$, the states contain $2n$ domains with opposite radial magnetization (see Fig.~\ref{fig.spinfig_numsol}), and as the radius increases, the DWs make up a vanishingly small part of the system. Consequently, the energy dependence becomes
\begin{equation}
    E_n = -\mathcal{K}_0 + \frac{2|n|E_\mathrm{DW}}{2\pi R} = -\mathcal{K}_0 + \frac{|n|E_\mathrm{DW}}{\pi}\kappa, 
\end{equation}
where $E_\mathrm{DW}$ is the DW energy. The dashed red line in Fig.~\ref{fig.E_of_curv_numsol} shows a linear fit of the energy at small curvatures for $n=1$, and the slope of the line gives $E_\mathrm{DW}=22.3~\mathrm{meV/\AA} $, in good agreement with the value directly calculated from the DW solution for the flat monolayer.

Fig.~\ref{fig.spinfig_numsol} shows some example illustrations of the magnetization of the numerical solutions of Eq.~(\ref{eq.sineGord_curv}), for different values of $n$ and $R$. At the smaller $R=10~\mathrm{\AA}$, where the $n=1$ state is lowest in energy, it essentially takes the form of the "perpendicular" magnetization state, or equivalently the $n=1$ cycloid, showing only small deviations of around $1^\circ$ from that configuration. However, at larger radii, the solutions for non-zero $n$ distinctly deviate from the $n$-cycloids, and instead take the form of $2n$ domains, magnetized along $\pm \hat{r}$.
\begin{figure}
	\centering
	\includegraphics[width=0.58\textwidth]{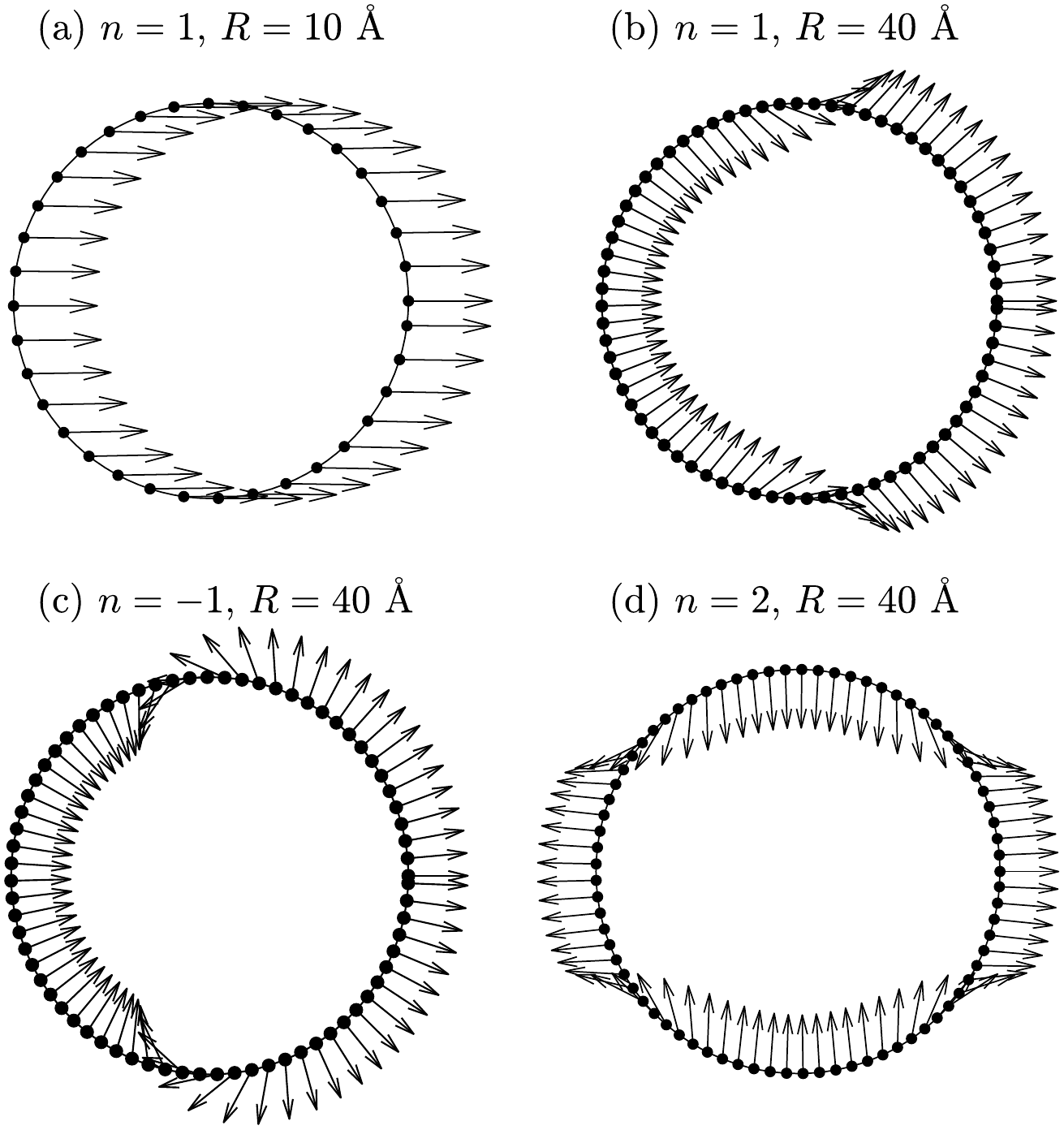}
	\caption{Illustrations of the magnetization of the stationary states of the continuum energy functional for the curved CrI$_3$ nanotubes, for different values of $n$ and $R$.}
	\label{fig.spinfig_numsol}
\end{figure}

}

\subsection{Non-Constrained Magnetic Calculations} 

\begin{figure}
	\centering
	\includegraphics[width=0.84\textwidth]{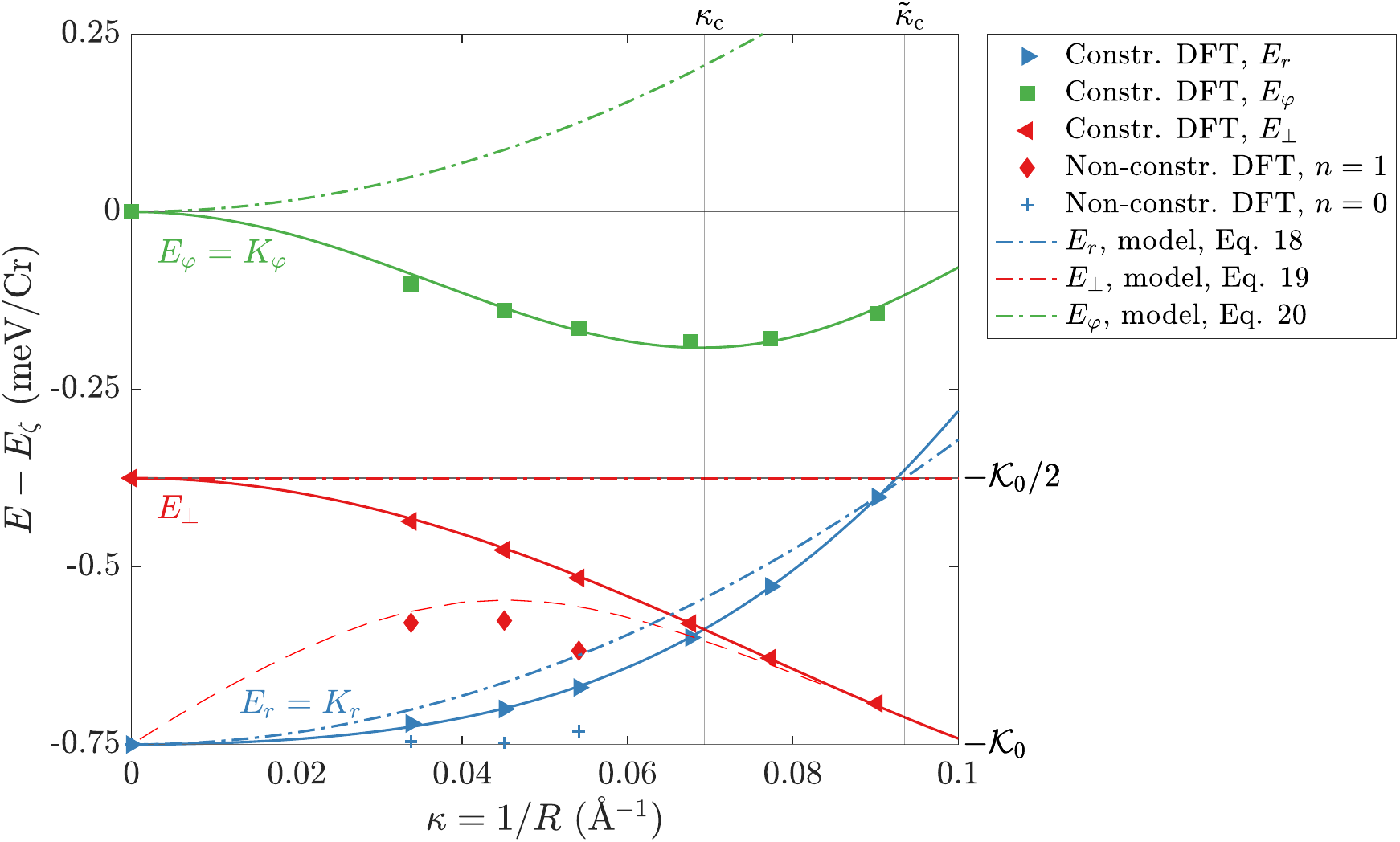}
	\caption{Energies of magnetic states, relative to $E_\zeta$, as function of curvature in CrI$_3$ nanotubes, using constrained or non-constrained, non-collinear magnetic DFT calculations, with SOC. Additionally, the energies predicted by the model in Eq.~(\ref{eq.Econt}) are shown as dash-dotted lines. The red dashed line shows the stationary state for $n=1$ of our continuum model.}
	\label{fig.compare_constr_vs_nonconstr}
\end{figure}

The DFT calculations discussed in the main text, and used for the fitting procedure discussed above, were done using a constraint of $\Lambda=5~\mathrm{eV}$ to keep the Cr spins in the directions defining each magnetic state considered. To check that the results are not significantly influenced by the constraint, we did some of the calculations also with $\Lambda=10~\mathrm{eV}$, confirming a negligible change to the energies. Next, we also performed calculations without constraints, but initializing the calculations with different spin configurations and allowing them to relax. The results for calculations initialized with radial magnetization ($n=0$) and the stationary states of the continuum model for $n=1$ are shown for the large radius nanotubes in Fig.~\ref{fig.compare_constr_vs_nonconstr}. The energies of the non-constrained calculations are lower compared to the constrained calculations, as is expected with the additional degrees of freedom. The spins of the non-constrained calculations have additional small components along the $\zeta$-axis. This additional level of complexity will be further analyzed in future work. However, the angle between the spins and the $(r,\varphi)$-plane remains of order $10^{-2}$ and the effect decreases with decreasing curvatures, whereby the dominant effects in the small curvature limit are captured by the continuum model in Eq.~\ref{eq.soc_cont_mod}. 

\subsection{Comparison to Other Continuum Models} 

Earlier work based on phenomenological continuum models~\cite{PhysRevLett.112.257203,PhysRevLett.114.197204,Sheka_2015,Streubel_2016} described curvilinear magnetism in 1D and 2D using a model considering two terms: one for the isotropic exchange stiffness and one anisotropy term. The energy in the 2D case is 
\begin{equation}\label{eq.Econt}
    E = \int \left[ A(\nabla \mathbf{m} )^2 - K_0 (\hat{\mathbf{m}} \cdot \hat{\mathbf{n}})^2 \right] \mathrm{d} S ,
\end{equation}
with magnetization density $\mathbf{m}$,  integration over the surface $S$ with normal $\hat{\mathbf{n}}$, and curvature-independent spin stiffness $A$ and anisotropy constant $K_0$. In this model, the effect of SOC is incorporated in the magnetic anisotropy of the second term. Based on Eq.~(\ref{eq.Econt}), the energy densities (per surface area) of the three magnetic states, discussed in the main text, are 
\begin{align} \label{eq.modelE}
    & E_r = -K_0 + A\kappa^2 \\ 
    & E_\perp = -K_0 / 2 \label{eq.modelE2} \\
    & E_\varphi =  A\kappa^2. 
     \label{eq.modelE3}
\end{align}
Using the calculated values for CrI$_3$, $A = \overline{A}_0 = 49.9$ (meV\AA$^{2}$/ Cr) and $K_0 = 0.75$ meV/Cr, in Eq.~(\ref{eq.modelE})-(\ref{eq.modelE3}), leads to the dash-dotted curves in Fig.~\ref{fig.compare_constr_vs_nonconstr}. Comparing these to the DFT results reveals clear qualitative and quantitative differences. For $E_\varphi$, Eq.~(\ref{eq.Econt}) predicts an increase in energy with curvature, relative to the axial magnetization, whereas the DFT calculations show that $E_\varphi$ decreases. In a material with negative $K_0$, favoring in-surface magnetization, this would lead to different predictions for the magnetic ground state of the curved material. 

Also in $E_\perp$, a qualitatively different behavior is found; while Eq.~(\ref{eq.modelE2}) predicts a curvature-independent value, the DFT calculations reveal that $E_\perp$ decreases with increasing curvature.

In the case of $E_r$, the model results and the DFT data agree that the energy increases with curvature. However, the leading order (quadratic) coefficient is sharply different, with the one for the dash-dotted line being $\overline{A}_0 = 49.9$ meV\AA$^{2}$/ Cr, and that of the solid line $K_{r,2} = 20.8$ meV\AA$^{2}$/ Cr. 

Regardless the qualitative differences in the predictions of Eq.~(\ref{eq.Econt}) and the DFT data, the crossover from the radial to the perpendicular magnetization state is observed in both cases (the red and blue lines cross each other, both for the dash-dotted and solid lines), although at different curvatures; the crossover is found to be $\kappa=0.070~\mathrm{\AA}$ from the DFT calculations and $\kappa=0.094~\mathrm{\AA}$ in the model of Eq.~(\ref{eq.Econt}). This indicates that the essential physics to describe the magnetic crossover is the competition between the usual magnetic anisotropy, occurring already without curvature due to SOC, and effective anisotropy or DMI terms emerging from the isotropic exchange interaction, as a result of curvature. Nevertheless, the complete first principles treatment, including SOC, leads to important alterations to the curvature dependence of the relevant terms in the magnetic energy, that may lead to distinctly different magnetic ground states in other materials. 

\subsection{Relation to Atomistic Spin Hamiltonians}

{

Beyond a continuum description, the atomistic spin Hamiltonian relevant to describe the magnetism of CrI$_3$ is~\cite{Xu2018,PhysRevB.102.115162} 
\begin{equation}\label{eq.Heis}
    H = -\frac{1}{2} \sum_{i,j}\sum_{\alpha,\beta} J_{ij}^{\alpha \beta} S_i^\alpha  S_j^\beta  - K \sum_i (S_i^z)^2,
\end{equation}
where $S_i^\alpha$ is Cartesian component $\alpha$ of the normalized Cr spin at site $i$, $K$ is the single ion anisotropy (SIA) constant and $J_{ij}^{\alpha \beta}$ is the $3\times 3$ exchange tensor connecting spins at sites $i$ and $j$. First-principles estimates of exchange parameters based on the energy-mapping four-state method~\cite{Xu2018} and magnetic force theorem~\cite{PhysRevB.102.115162} revealed significant anisotropy of the nearest-neighbor (NN) exchange tensor, while non-negligible isotropic exchange interactions up to third nearest neighbors have also been predicted~\cite{PhysRevB.102.115162, Besbes2019,Kashin_2020}. The local symmetry of the NN Cr-Cr bond reduces the number of independent components of the full exchange tensor from nine to four, and imposes the antisymmetric component (related to DMI) to be zero. In a local Cartesian reference frame with $y$ axis parallel to the Cr-Cr bond vector, the NN exchange tensor of the atomistic spin model reads:
\begin{equation}\label{Jtensor}
\mathbf{J}_{1} = \left(
\begin{array}{ccc}
J^{xx} & 0 & J^{xz} \\
0 & J^{yy} & 0 \\
J^{xz} & 0 & J^{zz}.
\end{array}
\right)
\end{equation}
The exchange tensors for the other NN bonds can then be easily obtained by transforming Eq. (\ref{Jtensor}) using the symmetry operations bringing one bond to another. In order to better identify the role played by the anisotropy of the exchange tensor, it is customary to separate isotropic and anisotropic parts as $J^{\alpha\beta}=J_{\mathrm{iso}}\delta_{\alpha,\beta}+J_{\mathrm{aniso}}^{\alpha\beta}$, where $J_{\mathrm{iso}}=\sum_\alpha J^{\alpha\alpha}/3$ and $J_{\mathrm{aniso}}^{\alpha\beta}=(J^{\alpha\beta}+J^{\beta\alpha})/2-J_{\mathrm{iso}}$.

It is straightforward to show that the anisotropy of the exchange interaction contributes to the total magnetic anisotropy energy. Indeed, for a flat monolayer the total magnetic anisotropy (per Cr), defined as the difference for in-plane and out-of-plane magnetization, is 
\begin{equation}
    \mathcal{K}_0 = K + \frac{3}{4}\left(2J_{\mathrm{aniso}}^{zz}-J_{\mathrm{aniso}}^{xx}-J_{\mathrm{aniso}}^{yy} \right) \equiv K+\frac{9}{4}J_{\mathrm{aniso}}^{zz} 
\end{equation}
where the last equality stems from the fact that the anisotropic exchange tensor $J_{\mathrm{aniso}}^{\alpha\beta}$ is traceless by definition. The total magnetic anisotropy originates not only from a SIA term but also from anisotropic exchange interactions. 
Using the atomistic model parameters estimated for the flat CrI$_3$ monolayer in Ref. \cite{Xu2018}, $K=0.59~\mathrm{meV}$ and $9 J_{\mathrm{aniso}}^{zz}/4=0.41~\mathrm{meV}$, one finds that roughly 40\% of the magnetic anisotropy arises from anisotropic exchange. 

The link between the atomistic model Eq. (\ref{eq.Heis}) and the continuum energy functional considered in this work is provided by assuming that magnetic moments rotate on a length scale that is much larger than the interatomic distance. This assumption formally allows to introduce a continuous normalized function $\mathbf{m}(\mathbf{R_i})=\mathbf{S}_i$, where $\mathbf{R}_i$ is the atomic position of the magnetic ion, and to use a Taylor expansion for the magnetization $\mathbf{m}(\mathbf{R}_j)$ around the atomic site $\mathbf{R}_i$. Using this standard procedure,
the spin stiffness tensor $\mathcal{A}$ can be expressed in terms of the isotropic exchange interactions $J_{ij}$ as~\cite{PhysRevB.100.214406}
\begin{equation}\label{stiff}
    \mathcal{A} = \frac{1}{4}\sum_{i\neq j} J_{ij} \mathbf{R}_{ij} \otimes \mathbf{R}_{ij},
\end{equation}
where $\mathbf{R}_{ij}=\mathbf{R}_i-\mathbf{R}_j$ is the separation vector between sites $i$ and $j$. Considering the flat CrI$_3$ monolayer with hexagonal symmetry and an isotropic NN exchange $J_{1,\mathrm{iso}}  = 4.84~\mathrm{meV}$ with NN Cr-Cr distance of $d=3.86~\mathrm{\AA}$, this leads to $A_0 = \mathcal{A}_{xx} = \mathcal{A}_{yy} = \frac{3}{8}J_{1,\mathrm{iso}}d^2 = 27.0~\mathrm{meV \AA^2 / Cr}$, that is smaller than the values of $42.9~\mathrm{meV\AA^{2}/ Cr}$ with SOC, or $40.7~\mathrm{meV\AA^{2}/ Cr}$ without SOC, found from DFT calculations of spin spiral energies here. Such disagreement is expected since only the NN exchange interactions was included in the calculation. Indeed, inclusion of up to third NN exchange interactions in Eq. (\ref{stiff}) modifies the spin stiffness of the flat CrI$_3$ monolayer as $A_0 =\frac{a_0^2}{8}\left(J_{1,\mathrm{iso}}+6J_{2,\mathrm{iso}}+4J_{3,\mathrm{iso}} \right)$, where $a_0=\sqrt{3}d$ is the lattice constant. Using published estimates of isotropic exchanges of CrI$_3$ obtained by different approaches~\cite{PhysRevB.102.115162,Besbes2019,Kashin_2020}, the spin stiffness is found to range between 35 and 50 $~\mathrm{meV\AA^{2}/ Cr}$.

The above analysis provides a link between the continuum energy functional considered here and microscopic, atomistic spin Hamiltonians. The discussion indicates that a complete microscopic understanding of the effect of curvature requires calculating both SIA and the anisotropic exchange interactions, possibly also including further neighbour interactions, as functions of curvature. On general grounds, bending the monolayer would reduce the symmetry of local bonds, for instance breaking the inversion symmetry that prevents the onset of DMI in the flat case: one can also speculate that the strong dependence on curvature of $\mathcal{K}$ arises from the anisotropic, symmetric part of the exchange tensor (whose parameters are expected to depend on bond length and angle) that largely contributes to the magnetic anisotropy.

}

\bibliography{literature}{}
\bibliographystyle{apsrev4-1}